\documentclass[aps,prx,showpacs,floatfix,twocolumn,superscriptaddress,longbibliography]{revtex4-2}

\usepackage{float}
\usepackage{color}
\usepackage{bm}
\usepackage{amssymb}
\usepackage{todonotes}
\usepackage{verbatim}
\usepackage{soul}
\usepackage{glossaries}
\usepackage{sidecap}
\usepackage{hyperref}
\hypersetup{
    colorlinks=true,
    linkcolor=blue,
    filecolor=magenta,
    urlcolor=red,
    citecolor=blue,
}
\usepackage{orcidlink}

\def\Q{\ensuremath{\bm{Q}}}

\def\La438{La$_{4}$Ni$_{3}$O$_{8}$}
\def\Pr438{Pr$_{4}$Ni$_{3}$O$_{8}$}
\def\Ni112{\textit{R}NiO$_2$}
\def\LSCO{La$_{2-x}$Sr$_{x}$CuO$_{4}$}
\def\Bi2223{Bi$_2$Sr$_2$Ca$_2$Cu$_3$O$_{10+\delta}$ ($\delta=0.18$)}

\newacronym{CO}{CO}{charge order}
\newacronym{JDOS}{JDOS}{joint density of states}
\newacronym{BS}{BS}{bond-stretching}
\newacronym{RIXS}{RIXS}{resonant inelastic x-ray scattering}
\newacronym{REXS}{REXS}{resonant elastic x-ray scattering}
\newacronym{XAS}{XAS}{x-ray absorption spectroscopy}
\newacronym{EELS}{EELS}{electron energy loss spectroscopy}
\newacronym{EPC}{EPC}{electron-phonon coupling}
\newacronym{CDW}{CDW}{charge density wave}
\newacronym{SDW}{SDW}{spin density wave}
\newacronym{FWHM}{FWHM}{full-width at half-maximum}
\newacronym{HWHM}{HWHM}{half-width at half-maximum}
\newacronym{INS}{INS}{inelastic neutron scattering}
\newacronym{DFT}{DFT}{density functional theory}
\newacronym{GGA}{GGA}{generalized gradient approximation}
\newacronym{UHB}{UHB}{upper Hubbard band}
\newacronym{ZSA}{ZSA}{Zaanen-Sawatzky-Allen}
\newacronym{ZRS}{ZRS}{Zhang-Rice singlet}
\newacronym{ED}{ED}{exact diagonalization}
\newacronym{CEF}{CEF}{crystal electric field}
\newacronym{2D}{2D}{two-dimensional}
\newacronym{3D}{3D}{three-dimensional}
\newacronym{TM}{TM}{transition-metal}
\newacronym{LDA}{LDA}{local density approximation}
\newacronym{DMFT}{DMFT}{dynamical mean field theory}
\newacronym{NSLS-II}{NSLS-II}{National Synchrotron Light Source II}
\newacronym{RPA}{RPA}{random phase approximation}
\newacronym{ARPES}{ARPES}{angle-resolved photoemission spectroscopy}
\newacronym{RP}{RP}{Ruddlesden-Popper}

\begin{document}

\title{Observation of correlated plasmons in low-valence nickelates}

\author{Y.~Shen\orcidlink{0000-0003-4697-4719}}
\email[]{yshen@iphy.ac.cn}
\affiliation{Condensed Matter Physics and Materials Science Department, Brookhaven National Laboratory, Upton, New York 11973, USA}
\email{Present address: Beijing National Laboratory for Condensed Matter Physics, Institute of Physics, Chinese Academy of Sciences, Beijing 100190, China}

\author{W.~He\orcidlink{0000-0003-3522-3899}}
\affiliation{Condensed Matter Physics and Materials Science Department, Brookhaven National Laboratory, Upton, New York 11973, USA}
\affiliation{Stanford Institute for Materials and Energy Sciences, SLAC National Accelerator Laboratory, Menlo Park, CA 94025, USA}
\author{J.~Sears\orcidlink{0000-0001-6524-8953}}
\affiliation{Condensed Matter Physics and Materials Science Department, Brookhaven National Laboratory, Upton, New York 11973, USA}

\author{Xuefei~Guo\orcidlink{0000-0003-1088-6039}}
\author{Xiangpeng~Luo\orcidlink{0000-0002-5471-053X}}
\author{A.~Roll\orcidlink{0009-0007-6140-0242}}  
\affiliation{Condensed Matter Physics and Materials Science Department, Brookhaven National Laboratory, Upton, New York 11973, USA}

\author{J.~Li\orcidlink{0000-0002-7153-6123}}
\author{J.~Pelliciari\orcidlink{0000-0003-1508-7746}}
\affiliation{National Synchrotron Light Source II, Brookhaven National Laboratory, Upton, New York 11973, USA}

\author{Xi~He\orcidlink{0000-0001-6603-2388}}
\affiliation{Condensed Matter Physics and Materials Science Department, Brookhaven National Laboratory, Upton, New York 11973, USA}

\author{I.~Bo\v{z}ovi\'{c}\orcidlink{0000-0001-6400-7461}}
\email{Shanghai Advanced Research in Physical Sciences (SHARPS), Pudong, Shanghai 201203, China}
\affiliation{Condensed Matter Physics and Materials Science Department, Brookhaven National Laboratory, Upton, New York 11973, USA}
\affiliation{Department of Chemistry, Yale University, New Haven, Connecticut 06520, USA}

\author{Junjie~Zhang\orcidlink{0000-0002-5561-1330}}
\affiliation{Materials Science Division, Argonne National Laboratory, Lemont, Illinois 60439, USA}
\affiliation{State Key Laboratory of Crystal Materials, Institute of Crystal Materials, Shandong University, Jinan, Shandong 250100, China}

\author{J.~F.~Mitchell\orcidlink{0000-0002-8416-6424}}
\affiliation{Materials Science Division, Argonne National Laboratory, Lemont, Illinois 60439, USA}

\author{V.~Bisogni\orcidlink{0000-0002-7399-9930}}
\affiliation{National Synchrotron Light Source II, Brookhaven National Laboratory, Upton, New York 11973, USA}

\author{M.~Mitrano\orcidlink{0000-0002-0102-0391}}
\affiliation{Department of Physics, Harvard University, Cambridge, Massachusetts 02138, USA}

\author{S.~Johnston\orcidlink{0000-0002-2343-0113}}
\affiliation{Department of Physics and Astronomy, The University of Tennessee, Knoxville, Tennessee 37966, USA}
\affiliation{Institute of Advanced Materials and Manufacturing, The University of Tennessee, Knoxville, Tennessee 37996, USA\looseness=-1}

\author{M.~P.~M.~Dean\orcidlink{0000-0001-5139-3543}}
\email[]{mdean@bnl.gov}
\affiliation{Condensed Matter Physics and Materials Science Department, Brookhaven National Laboratory, Upton, New York 11973, USA}
\affiliation{Department of Physics and Astronomy, The University of Tennessee, Knoxville, Tennessee 37966, USA}

\date{\today}

\begin{abstract}
The discovery of nickelate superconductors has opened a new arena for studying the behavior of correlated electron liquids that give rise to unconventional superconductivity. While critical information about a material's charge dynamics is encoded in its plasmons, collective modes of the electron gas, these excitations have not yet been observed in nickelate materials. Here, we use \acrfull*{RIXS} to detect plasmons in the metallic, low-valence nickelate \Pr438{}. Although qualitatively similar to those in cuprates, the nickelate plasmons are more heavily damped and have a lower velocity than those in a cuprate at comparable doping, which we attribute to reduced electronic hopping and enhanced screening of the long-range Coulomb interactions. Furthermore, the plasmons in \Pr438{} soften with increasing temperature, in contrast to the cuprate, where plasmons remain at nearly fixed energy but become more strongly damped. Taken together, these results reveal a distinct charge-screening landscape in nickelates and place quantitative constraints on analogies to cuprates.
\end{abstract}

\maketitle

\section{Introduction}

The discovery of superconductivity in nickelates crowns decades of efforts to realize cuprate-like physics in materials that do not contain copper \cite{Crespin1983reduced, Anisimov1999electronic, Norman2020entering, Zhang2021review, Wang2024Experimental}. Two distinct subclasses of nickelates have been identified thus far: low-valence nickelates with formula $R_{n+1}$Ni$_n$O$_{2n+2}$ (where $R$ is a rare earth and $n$ denotes the number of neighboring layers)~\cite{Li2019superconductivity,Pan2021super,Chow2025Bulk,Yang2025Enhanced,Yan2025Superconductivity}, and \gls*{RP} phase nickelates with formula $R_{n+1}$Ni$_n$O$_{3n+1}$~\cite{Sun2023Signatures,Zhu2024Superconductivity,Ko2025Signatures,Zhou2025Ambientpressure,Shi2025Superconductivity}. Whereas the \gls*{RP} phases exhibit disparate band topology, the low-valence nickelates have band structures that share notable similarities with those of the cuprates, such as comparable electron counts~\cite{Nomura2022Superconductivity,Lee2023Linearintemperature,Wang2024Experimental, Hepting2021soft}. However, low-valence nickelates possess a larger charge-transfer energy compared to cuprates, placing them further from the charge-transfer regime~\cite{ZSA1985} and correspondingly reducing their magnetic exchange interactions~\cite{Hepting2020electronic, Lu2021magnetic, Lin2021strong, Shen2022Role, Shen2023electronic, Fabbris2023resonant, Norman2023orbital}. 

\begin{figure*}
\includegraphics{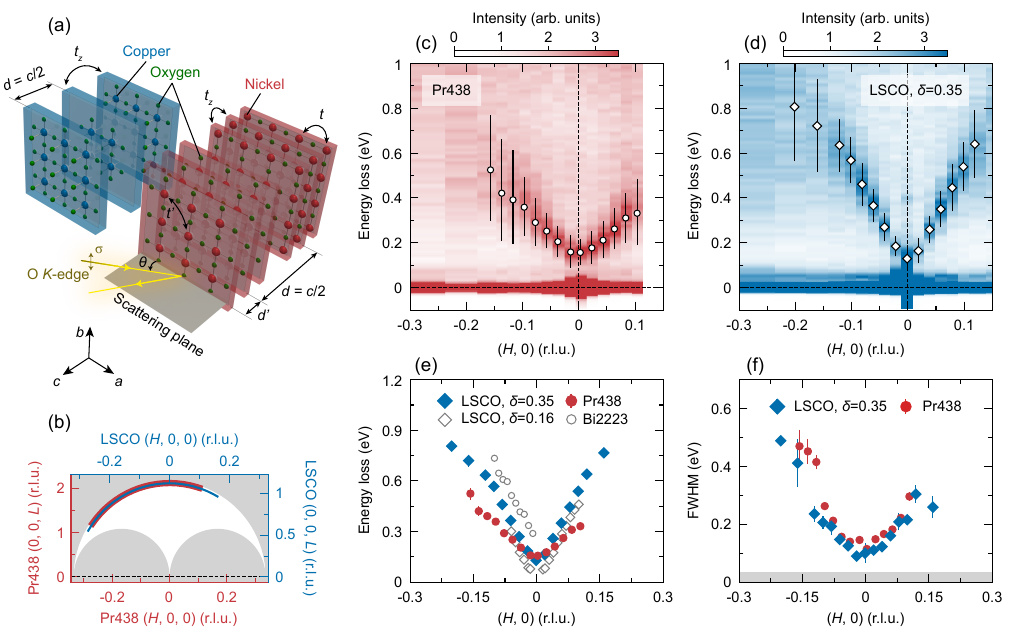}
\caption{Dispersive plasmons in \Pr438{}. (a), Schematic of the crystal structures of \Pr438{} (Pr438, red) and \LSCO{} (LSCO, blue), along with the corresponding structural and electronic hopping parameters. For the single layer model, $d$ is the interlayer distance, and $t_z$ denotes the interlayer electronic hopping integral. For the trilayer model, $d$ and $d^\prime$ are the interlayer and intra-trilayer distances, respectively. $t_z$ denotes the intra-trilayer hopping, while inter-trilayer hopping is neglected, as it is expected to be much smaller than the intra-trilayer term. $t$ and $t^\prime$ are the nearest- and next-nearest-neighbor in-plane hopping for both \Pr438{} and \LSCO{}. (b), Reciprocal space trajectories of the \acrfull*{RIXS} measurements of the in-plane plasmon dispersions, which were performed at a fixed scattering angle $2\Theta$. The values presented are in r.l.u.\ (reciprocal lattice units). (c),(d), RIXS intensity maps of \Pr438{} and \LSCO{} ($\delta=0.35$), respectively, collected at 40~K, exhibiting dispersive plasmons with momenta primarily along the ($H$, 0) direction, following the \Q{} trajectories presented in (b). The markers indicate plasmon peak positions extracted from fitting (See Ref.~\cite{supp} Sec.~S2). (e),(f) Summary of the in-plane plasmon dispersions, showing the peak positions and \acrfull*{FWHM} that we obtain from fitting. Data for \LSCO{} ($\delta=0.16$) and Bi$_2$Sr$_2$Ca$_2$Cu$_3$O$_{10+x}$ (Bi2223, $\delta=0.18$) are reproduced from Ref.~\cite{Nag2020detection} and Ref.~\cite{Nakata2025Outofphasea}, respectively. The shaded area in (f) represents the quasi-elastic regime, determined by the energy resolution of $\sim30$~meV. The full extent of the error bars in (c),(d) denotes the \gls*{FWHM} of plasmon peaks, whereas all other error bars are 1-$\sigma$ confidence intervals evaluated from the fitting.}
\label{fig:Hdisp}
\end{figure*}

Although much is now known about nickelate band structures, it remains crucial to understand the dynamical properties of their normal state. These depend sensitively on the strength of the electron interactions and in particular screened Coulomb interactions, which are difficult to quantify from band structure measurements. This gap presents a compelling opportunity: recent studies indicate that attractive extended-Hubbard interactions---likely phonon-mediated---can promote superconductivity, but only if they are strong enough to overcome the long-range Coulomb repulsion \cite{Chen2021anomalously, Jiang2022enhancing, Padma2025beyond, Scheie2025cooper}. Quantifying the magnitude of this Coulomb interaction is therefore essential. These Coulomb interactions manifest in plasmon excitations, which are the collective charge-density oscillations of low-energy electrons~\cite{Bozovic1900plasmons,Foussats2004Large,Greco2016Plasmon,Zinni2023Lowenergy}. Plasmons have been studied extensively in cuprates \cite{Bozovic1900plasmons, Abbamonte2025Collective, Mitrano2024Exploring,Hepting2018Threedimensional,Lin2020Doping,Nag2020detection,Hepting2022Gapped,Singh2022Acoustic,Hepting2023Evolution,Bejas2024Plasmon,Nag2024Impacta,Nakata2025Outofphasea}, but they are yet to be observed in nickelates. Filling this gap is imperative to directly compare screened interactions in nickelates and cuprates, and to identify universal features that are key to the formation of unconventional superconductivity.

Plasmons are traditionally measured using techniques such as optics and \gls*{EELS}~\cite{Nucker1989plasmons, Nucker1991long, Levallois2016TemperatureDependent, Abbamonte2025Collective}, but \gls*{RIXS} has emerged as an alternative~\cite{Mitrano2024Exploring, Hepting2018Threedimensional, Lin2020Doping, Nag2020detection, Hepting2022Gapped, Singh2022Acoustic, Hepting2023Evolution, Bejas2024Plasmon, Nag2024Impacta, Nakata2025Outofphasea}. It is especially useful for accessing the out-of-plane momentum dependence \cite{Hepting2018Threedimensional, Nag2020detection}, which is important because collective charge excitations in layered materials often disperse strongly along this direction, and the lowest-energy modes typically occur at finite momentum.

\begin{figure*}
\includegraphics{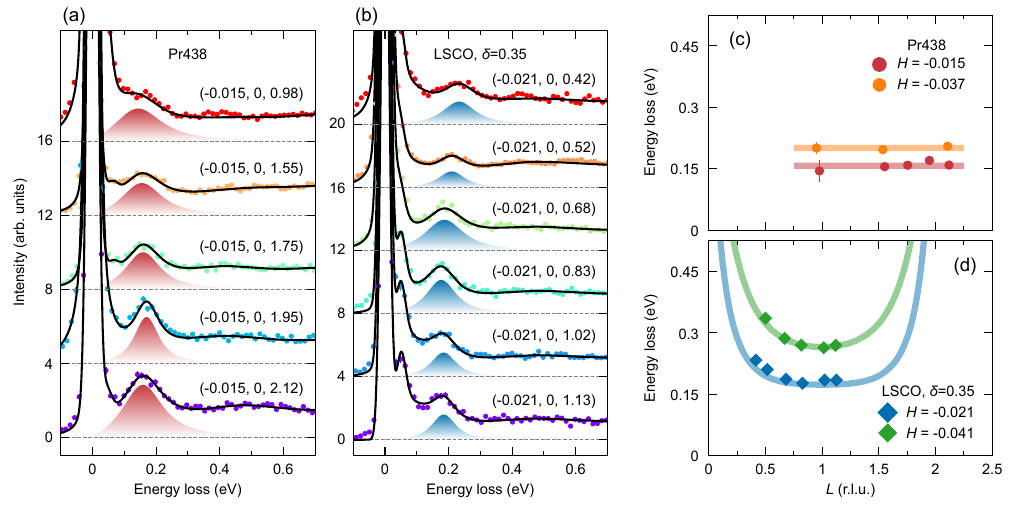}
\caption{Out-of-plane plasmon dispersions. (a),(b), RIXS spectra of various \Q{} along the out-of-plane $L$ direction for \Pr438{} and \LSCO{} ($\delta=0.35$), respectively. The shaded peak profiles represent the fitted plasmon contributions, and the solid black lines indicate the fitting results with all components accounted for, including the quasi-elastic line, low-energy phonons, and high-energy bimagnons combined with a continuous charge background. (c), Fitted plasmon peak positions of \Pr438{}. (d), Fitted plasmon peak positions of \LSCO{}. The solid lines are guides to the eye.}
\label{fig:Ldisp}
\end{figure*}

Here, we use O $K$-edge \gls*{RIXS} to study the plasmon excitations in single crystalline effectively overdoped low-valence nickelate \Pr438{}, which we compare with overdoped cuprate \LSCO{} with a similar effective hole-doping level. The nickelate exhibits well-defined dispersive plasmons at small in-plane momenta, which become overdamped well before crossing the particle-hole continuum inferred from the band structure. The plasmons in \Pr438{} exhibit a reduced velocity compared to cuprates, indicating suppressed electronic hopping, and a broader linewidth, suggestive of more strongly screened long-range Coulomb interactions. Our \gls*{RPA} calculations corroborate these observations, which also account for the lack of clear out-of-plane plasmon dispersion in \Pr438{}. We further observe that the plasmons in \Pr438{} soften with increasing temperature, contrasting sharply with \LSCO{}, where they maintain their excitation energy but experience increased damping at elevated temperatures. Our findings reveal fundamental differences in charge dynamics between nickelate and cuprate superconductors, providing critical experimental constraints for identifying parameters essential to unconventional superconductivity.

\section{Observation of dispersing plasmons in nickelates}

We choose \Pr438{} as our reference low-valence nickelate [Fig.~\ref{fig:Hdisp}(a)] for this study and prepared samples via floating-zone growth, followed by topotactic reduction (see Sec.~\ref{sec:synthesis}). This material possesses a Ni $3d^{9-1/3}$ valence state, corresponding to $\delta=1/3$ hole doping with respect to the $3d^9$ parent state, placing it within the overdoped non-superconducting regime~\cite{Zhang2017large}. Distinct from its counterpart \La438{}, which undergoes a semiconductor-insulator transition accompanied by intertwined charge and spin order~\cite{Zhang2016stacked,Zhang2019stripe}, \Pr438{} remains metallic down to 2~K with no identifiable phase transitions~\cite{Zhang2017large}. The family of low-valence multi-layer nickelates has been shown to superconduct with a maximum $T_c$ of 12.9~K \cite{Pan2021super,Pan2026Superconducting}.  We compare \Pr438{} with the overdoped single-layer cuprate \LSCO{} since high-quality overdoped trilayer cuprate single crystals are currently unavailable. While doping-induced plasmon renormalization in these strongly correlated materials is not yet fully understood, layer-structure effects can be directly probed via out-of-plane plasmon dispersion measurements and incorporated into the modeling, enabling insightful comparisons. Both materials reside in the overdoped, non-superconducting regime. Studying these materials provides insight into the normal state from which superconductivity emerges and thus can provide crucial information about the mechanisms underlying superconductivity.


We begin by identifying the plasmon and mapping out its momentum dependence at 40~K. O $K$-edge \gls*{RIXS} data were collected as detailed in Sec.~\ref{sec:RIXS}. Figure~\ref{fig:Hdisp}(c),(d) display the \gls*{RIXS} energy-momentum maps for \Pr438{} and \LSCO{} ($\delta=0.35$), respectively, along the ($H$, 0) direction. The resonant condition for these measurements was determined by scanning the incident x-ray energy through the O $K$-edge pre-peak and identifying the energy that maximizes the intensity of the low-energy electronic excitations (see Ref.~\cite{supp} Sec.~S1). Both materials exhibit well-defined dispersive modes below 1~eV energy loss, which are widely recognized as plasmons in cuprate superconductors \cite{Hepting2018Threedimensional,Lin2020Doping,Nag2020detection,Hepting2022Gapped,Singh2022Acoustic,Hepting2023Evolution,Bejas2024Plasmon,Nag2024Impacta,Nakata2025Outofphasea}, as will be further verified by calculations presented later in this work. In electron-doped cuprates, as the doped electrons primarily populate the upper Hubbard band of Cu $3d$ orbitals, the charge near the Fermi level exhibits dominant Cu $3d$ character. Consequently, plasmons are prominently observed at the Cu $L$ edge in \gls*{RIXS}~\cite{Hepting2018Threedimensional,Nag2024Impacta}. In contrast, the doped carriers in hole-doped cuprates tend to occupy the ligand oxygen orbitals, making plasmons barely detectable at the Cu $L$ edge but substantial at the O $K$ edge~\cite{Nag2020detection,Nag2024Impacta}. The clear observation of plasmon excitations in \Pr438{} therefore implies an appreciable hole density on oxygen sites, supporting the observed mixed charge-transfer-Mott-Hubbard character of low-valence nickelates~\cite{Shen2022Role,Shen2023electronic}.

\begin{figure*}
\includegraphics{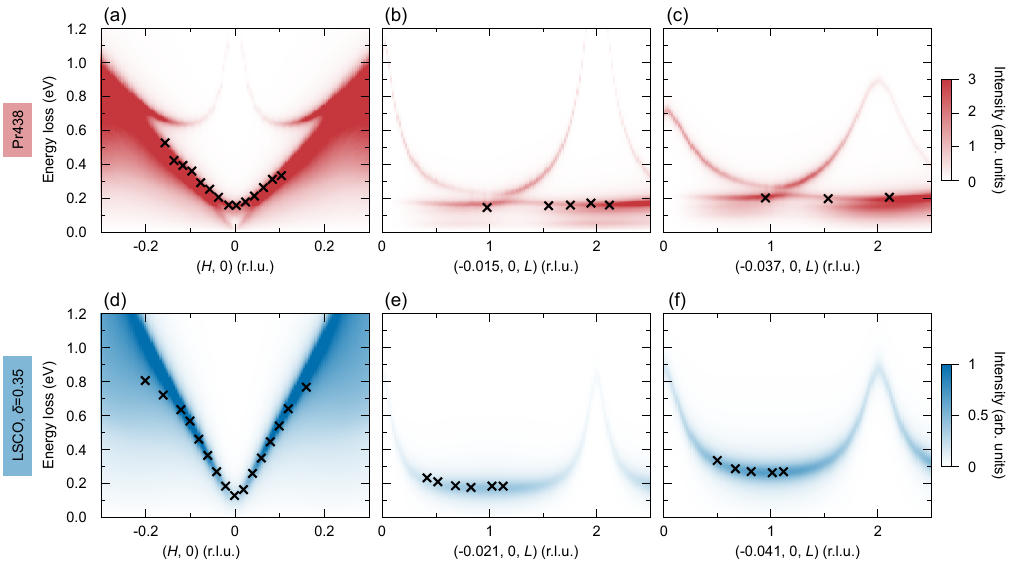}
\caption{Theoretical description of the dispersing plasmons. (a)--(c), Simulations of plasmons in \Pr438{} based on \gls*{RPA} calculations of the dynamical susceptibility applied to a trilayer model, as described in the main text. (d)--(f), \gls*{RPA} simulations of plasmons in \LSCO{} ($\delta=0.35$) applied to a single-layer model. The crosses represent plasmon peak positions derived from fitting the experimental data. In (a) and (d), the simulations followed the experimental \Q{} trajectory as shown in Fig.~\ref{fig:Hdisp}(b), while in other panels, the in-plane momentum transfer $H$ was fixed at the indicated values.}
\label{fig:calc}
\end{figure*}

Compared to the cuprate, \Pr438{} exhibits plasmons with a lower energy scale and reduced dispersion slope [Fig.~\ref{fig:Hdisp}(e)], indicating a lower velocity ($1.2\pm0.1$~eV\AA{} in \Pr438{} versus $2.4\pm0.2$~eV\AA{} in \LSCO{}). This distinction persists for lower dopings (for example \LSCO{} $\delta=0.16$) and in trilayer cuprates, as seen in Fig.~\ref{fig:Hdisp}(e). Here, we define the plasmon velocity using the dispersion gradient of the observed low energy modes. For \Pr438{} and trilayer cuprates, only one collective mode can be observed, which does not show strong dependence on the out-of-plane momentum, as will be discussed subsequently, making the plasmon velocity an empirical parameter independent of specific modeling. Meanwhile, plasmons in both \Pr438{} and \LSCO{} are heavily damped across reciprocal space, evidenced by peak widths exceeding the energy resolution of 30~meV [Fig.~\ref{fig:Hdisp}(f)], with the damping of the nickelate plasmon systematically exceeding that of the cuprate. In both cases, the damping increases strongly with momentum and with the nickelate disappearing more quickly with momentum than the cuprate.

The plasmon velocity is determined primarily by electronic hopping, with only a modest enhancement from long-range Coulomb interactions. More specifically, with the long-range Coulomb interactions removed, the plasmon velocity for \LSCO{} is estimated to be approximately 1.8~eV~\AA{} (Fig.~S4), which remains significantly larger than that observed in \Pr438{}. Thus, the lower plasmon velocity in \Pr438{} reflects weakened electronic hopping. Regarding damping, it is driven primarily by suppressed long-range Coulomb interactions~\cite{Zinni2023Lowenergy}, which facilitate the decay of plasmons into the particle-hole continuum.

\begin{table*}
\caption{Parameters describing plasmons in the two materials. Here, $t$, $t^\prime$, and $t_z$ are electronic hopping integrals, as defined in Fig.~\ref{fig:Hdisp}(a), and $V_c$ is the long-range Coulomb interaction with its spatial anisotropy controlled by $\alpha$. Note that $\mu$ is the chemical potential while $\Delta\mu$ is the potential difference between the outer and inner layers of the trilayer model. Full details are provided in Sec.~\ref{sec:RPA1}\&\ref{sec:RPA2}.}
\begin{ruledtabular}
\begin{tabular}{cccccccccccc}
Material & $t$ (eV) & $t^\prime/t$ & $t^\prime$ (eV) & $t_z/t$ & $t_z$ (eV) & $\alpha$ & $V_c/t$ & $V_c$ (eV) & $\mu$ (eV) & $\Delta\mu/t$ & $\Delta\mu$ (eV) \\
\hline
LSCO ($\delta=0.35$) & 0.39 & -0.3 & -0.117 & 0.017 & 0.00663 & 3.5 & 15 & 5.85 & -0.51 & -- & -- \\
Pr438 & 0.21 & -0.3 & -0.063 & 0.04 & 0.0084 & -- & 6 & 1.26 & -0.28 & -0.4 & -0.084 \\
\end{tabular}
\end{ruledtabular}
\label{table:Parameters}
\end{table*}


A key characteristic of cuprate plasmons is their out-of-plane periodic dispersion highlighting the importance of interlayer coupling beyond the typically assumed \gls*{2D} framework for cuprates~\cite{Hepting2018Threedimensional, Nag2020detection}. Figure~\ref{fig:Ldisp} summarizes the out-of-plane dispersion. As expected, our \LSCO{} ($\delta=0.35$) \gls*{RIXS} data reveal clear dispersive plasmons along the out-of-plane direction [Fig.~\ref{fig:Ldisp}(b),(d)], matching earlier reports. In contrast, \Pr438{} exhibits no resolvable out-of-plane plasmon dispersion over a \Q{} range spanning the full anticipated period [Fig.~\ref{fig:Ldisp}(a),(c)]. Similar behavior was observed in a trilayer cuprate, which was attributed to out-of-phase plasmon oscillations that become undetectable by \gls*{RIXS}~\cite{Nakata2025Outofphasea}. Therefore, the lack of out-of-plane plasmon dispersion likely reflects the presence of three coupled planes in \Pr438{}, rather than indicating a fundamental difference in the interactions.

\section{Theoretical description of the dispersing plasmons}

To better understand the implications of our observations, and quantify changes in hopping and screening in these materials, we simulated the plasmons using \gls*{RPA} calculations of the charge dynamical susceptibility, which incorporates dynamically screened interactions arising from collective, mean-field-like density fluctuations. Further correlation effects beyond this approximation can be partly captured through parameter renormalization~\cite{Nag2024Impacta}. Such \gls*{RPA} simulations can therefore associate the differences in plasmon behavior between nickelates and cuprates with specific variations in the underlying interactions in the materials. During the simulations, most parameters were fixed based on established theoretical and experimental literature, leaving only three primary free parameters for each material, which were determined exclusively from fits to the plasmon dispersions observed in our \gls*{RIXS} measurements (see Secs.~\ref{sec:RPA1}~\&~\ref{sec:RPA2} for more details).

For \LSCO{}, we employ a single-layer model and consider anisotropic long-range Coulomb interactions and intralayer/interlayer electronic hopping (Methods). Satisfactory agreement occurs using the parameters presented in Table~\ref{table:Parameters}, which reproduce plasmon dispersions in both in-plane and out-of-plane directions [Fig.~\ref{fig:calc}(d)--(f)].

\begin{figure*}
\includegraphics{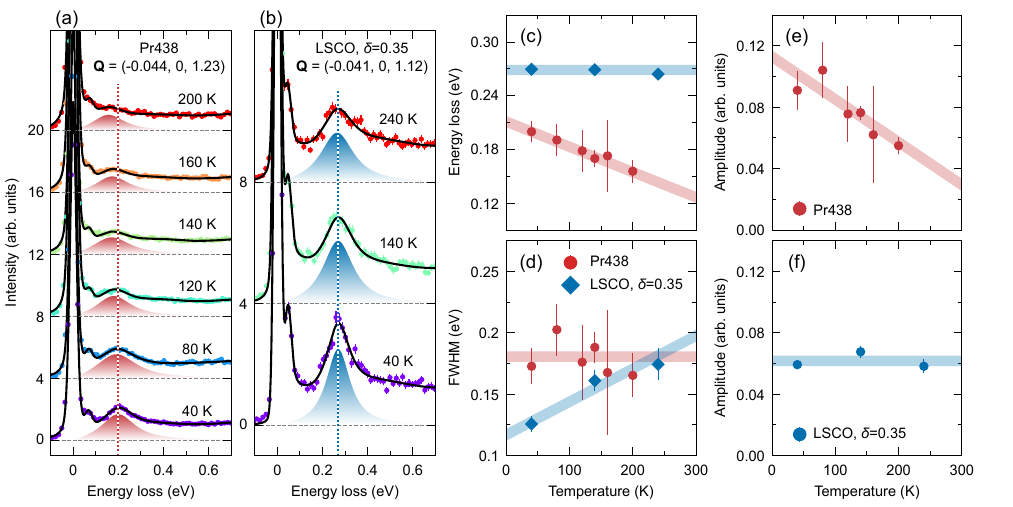}
\caption{Temperature dependence of the plasmons. (a),(b), Temperature dependent RIXS spectra for \Pr438{} and \LSCO{} ($\delta=0.35$), respectively, collected at the indicated \Q{} vectors and temperatures. The shaded areas indicate the fitted plasmon profiles, and the black lines denote the summation of different contributions. (c),(d), Temperature dependence of the plasmon peak positions and \gls*{FWHM}, respectively, obtained from the fitting. (e),(f), Fitted plasmon amplitudes for \Pr438{} and \LSCO{}, respectively. The dashed lines in (a),(b), and the bold lines in (c)--(f) are guides to the eye.}
\label{fig:Tdep}
\end{figure*}


For \Pr438{}, we construct a corresponding trilayer model, incorporating intra-trilayer hopping while neglecting inter-trilayer terms, as the former dominates (Methods). The simulations achieve reasonable agreement with experiment, as depicted in Fig.~\ref{fig:calc}(a)--(c), using the parameters presented in Table~\ref{table:Parameters}. Notably, \Pr438{} exhibits substantially reduced in-plane hopping and long-range Coulomb interaction $V_c$ compared to \LSCO{} ($\delta=0.35$), consistent with the reduced plasmon velocity of $1.2\pm0.1$~eV\AA{} compared to $2.4\pm0.2$~eV\AA{} in the cuprate [Figs.~\ref{fig:Hdisp}(c),(d)], linking the observed changes in plasmon velocity to changes in the underlying Hamiltonian.  Meanwhile, the intra-trilayer hopping in \Pr438{} is larger than the interlayer hopping in \LSCO{} due to its reduced interlayer distance, despite the absence of apical oxygens [Fig.~\ref{fig:Hdisp}(a)].

Although simulations of the charge dynamical susceptibility predict multiple plasmon modes [Fig.~\ref{fig:calc}(a)--(c)], only one mode is experimentally resolved. For most \Q{} vectors, the upper branches are predicted either to have intensities below the \gls*{RIXS} detection threshold ($L<0.4$) or to merge into the high-energy charge continuum ($L\sim2$), making them hard to distinguish from the background~\cite{Bejas2024Plasmon,Nakata2025Outofphasea,Yamase2025Theory,Sellati2025Ghost}. However, at specific \Q{} vectors like those in Fig.~\ref{fig:calc}(c) with $L\sim1.5$, \gls*{RPA} predicts that plasmon branches stay well-separated at low energies, with sufficient intensity for detection; yet they remain unobserved in the experiment.  Thus, alternative mechanisms need to be considered. One possibility is that matrix element effects suppress these modes in the \gls*{RIXS} process. More sophisticated multi-orbital calculations, taking into account the full Kramers-Heisenberg cross-section would be needed to address this possibility. Another more likely possibility is that these modes are obscured by correlations beyond those captured in standard \gls*{RPA} calculations. For example, self-energy broadening, which becomes increasingly pronounced as the quasiparticle energy shifts away from the Fermi level~\cite{Li2023Electronic}, can broaden the higher-energy plasmon branches significantly, exceeding observability, while the lower-energy modes remain coherent. This scenario is supported by \gls*{RPA} simulations incorporating phenomenological energy-dependent self-energy broadening (Ref.~\cite{supp} Fig.~S7).

To ensure our conclusions remain robust despite the limitations of \gls*{RPA}, we applied the same methodology to both materials and derived parameters exclusively from \gls*{RIXS} data. This approach ensures a consistent comparative framework, where any correlation effects unaccounted for by the \gls*{RPA} influence the analysis of both systems in a similar way. Consequently, even if absolute parameter values might shift slightly with more sophisticated models, the qualitative differences between the two materials should persist. Furthermore, the reduction in effective electronic hopping and long-range Coulomb interactions in low-valence nickelates is evident from a direct comparison between the trilayer \Pr438{} and single-layer \LSCO{}, alongside literature data for underdoped trilayer cuprates [Fig.~\ref{fig:Hdisp}(e),(f)], without reliance on any specific modeling.

\section{Temperature dependence}

Having established a qualitative picture of low-temperature plasmons, we now explore their response to thermal fluctuations. Figure~\ref{fig:Tdep} presents temperature-dependent \gls*{RIXS} measurements and the corresponding fitting results. For \LSCO{} ($\delta=0.35$), both the plasmon excitation energy and amplitude remain nearly constant while the profile broadens with increasing temperature, consistent with prior reports~\cite{Hepting2023Evolution}. In contrast, plasmons in \Pr438{} soften at higher temperatures with decreased intensity while maintaining similar peak widths (Fig.~\ref{fig:Tdep}).

The opposite trends of plasmon behavior in the cuprate and nickelate point to rich physics beyond those that can be interpreted by a simple self-energy effects. The lower plasmon velocity in \Pr438{} reflects greater electronic compressibility and thus an increased susceptibility to charge order in conditions where this instability is energetically favored. These novel plasmon effects are intriguing in light of prior evidence of incipient stripe order in \Pr438{} \cite{Huangfu2020shortrange, Lin2021strong, Hepting2021soft}. Since studies of the related material \La438{} show that stripe order modifies the material's electrical conductivity and consequently the plasmon screening \cite{Zhang2016stacked, Zhang2017large}, stripe fluctuations are a possible candidate for the temperature dependence of the plasmon. In such a scenario, an anomaly might be expected near the stripe ordering vector, \Q{}$_{\parallel}=(1/3, 1/3)$. However, since collective plasmon modes decay into the particle-hole continuum at large momenta, they are not observable at this \Q{} vector, precluding a definitive test of this hypothesis. Other factors, such as the multi-orbital nature of the low-energy electronic state, may also contribute to the temperature dependence. We note that typical temperature-induced renormalization would entail concurrent changes in both energy and linewidth.

\section{Implications of nickelate charge dynamics}

Our results provide the first measurements of dispersive plasmons in low-valence nickelate \Pr438{} and compare them with those in \LSCO{} with a similar doping level. The plasmon velocity, indicated by the dispersion slope in Fig.~\ref{fig:Hdisp}(e), is found to be smaller in \Pr438{} than \LSCO{}, implying reduced electronic hopping in the former. 
Prior experimental and theoretical work has shown that \Pr438{} shares similar electronic properties to other low valence nickelates, provided that they are compared at the same effective doping~\cite{Nica2020theoretical, Karp2020comparative, Lin2021strong, Shen2022Role, Shen2023electronic}, so this trend of suppressed electronic hopping and enhanced screening is likely to be a general property of low valence nickelates compared to the cuprates. 

At first glance, the trend in the plasmon velocities is perhaps surprising as the nominally Ni$^{1+}$ site in nickelates hosts a smaller nuclear charge than the Cu$^{2+}$ site in cuprates and so it would be expected to have more extended \gls*{TM} $d$-orbitals and \textit{larger} hopping. Indeed, prior works have found that the \gls*{TM}-O hopping values in the low-valence nickelates exhibit are larger, or at least similar to the values found in cuprates \cite{Shen2022Role, Nica2020theoretical}. Instead, the difference in the plasmon dispersion can be explained by the fact that the transfer energy is also larger in low-valence nickelates, so the effective \gls*{TM}-\gls*{TM} hopping bridged by oxygen is weaker. The more extended Ni orbitals are also likely to increase screening and have a role in the smaller long-range Coulomb interaction, which is almost a factor of five smaller in the nickelate compared to the cuprate. It is important to note that the hopping parameters extracted here are not bare quantities --- such as those obtained from \gls*{DFT} calculations --- but already incorporate (to some degree) renormalization effects arising from short-range Coulomb interactions. For instance, the bare nearest-neighbor hopping $t=0.43$~eV derived from tight-binding modeling of underdoped \LSCO{}~\cite{Markiewicz2005Oneband} is comparable to our \gls*{RPA}-extracted value (0.39~eV). In contrast, for \Pr438{}, the bare $t=0.39$~eV~\cite{Nica2020theoretical} is 86\% larger than the value extracted here (0.21~eV). This indicates that short-range correlation effects are stronger in low-valence nickelates than in cuprates, consistent with the conclusions of a recent \gls*{ARPES} study~\cite{Li2023Electronic}. These short-range correlation effects are distinct from the long-range ones, which are conversely weaker in low-valence nickelates.

Another notable feature of the plasmons observed here is that they are more heavily damped and disappear at smaller values of in-plane momentum compared to the cuprate. In a simple uncorrelated metal, plasmons become heavily damped only once their momentum crosses the particle-hole continuum. Indeed, the implications of the anomalous damping of cuprate plasmons has been discussed extensively \cite{Nag2020detection, Mitrano2018Anomalous, Husain2019crossover} and the microscopic origin of this damping remains controversial. Particular controversy has arisen from inconsistencies among different \gls*{EELS} measurements. Early results suggested that plasmons in cuprates could be understood within a conventional uncorrelated framework \cite{Nucker1989plasmons, Nucker1991long}, whereas later studies emphasized the crucial role of electronic correlations \cite{Mitrano2018Anomalous, Husain2019crossover}. Review articles have concluded that further investigation is required before a definitive comparison with \gls*{EELS} can be established \cite{Abbamonte2025Collective}. Consequently, we focus on comparisons with \gls*{RIXS}, which consistently shows that plasmons damp rapidly with increasing momentum \cite{Mitrano2024Exploring, Hepting2018Threedimensional, Lin2020Doping, Nag2020detection, Hepting2022Gapped, Singh2022Acoustic, Hepting2023Evolution, Bejas2024Plasmon, Nag2024Impacta, Nakata2025Outofphasea}. Possible explanations for the strong damping range from strange metal physics in which plasmons decay into a quantum critical continuum~\cite{Romero-Bermudez2019Anomalous}, or by interband transitions, which effectively extend the Landau damping regime~\cite{Paasch1970Influence}. We find that this anomalous plasmon damping is even stronger in nickelates, which can shed further light on this problem. The temperature dependence of the nickelate plasmon also appears to be anomalous, exhibiting a strong softening with increased temperature that may be connected to stripe correlations. While the present data are not sufficient to confirm or refute these interpretations, as the first report of plasmons in low-valence nickelates, this work should stimulate further theoretical efforts. These include approaches extending beyond the conventional Hubbard framework --- such as those incorporating long-range Coulomb interactions or modern holographic methods --- to better understand the electronic structure and mechanisms underlying unconventional superconductivity in nickelates.

Overall, our findings establish reduced electronic hopping and smaller, more strongly screened, long-range Coulomb interactions as defining features differentiating low-valence nickelates from cuprates. These differences could ultimately lead to a suppression of the superconducting energy scale \cite{Leggett1999Cuprate}, providing a possible explanation for the lower $T_c$ in low‑valence nickelates compared to cuprates. These results offer essential experimental benchmarks for contrasting low-valence nickelates and cuprates, guiding efforts toward realizing unconventional superconductivity with higher transition temperatures.

\begin{acknowledgments}

We thank Mike Norman, Hiroyuki Yamase, Andres Greco, Matias Bejas, and Peter Abbamonte for discussions. Work at Brookhaven and the University of Tennessee (RIXS measurements and the interpretation and model Hamiltonian calculations) was supported by the U.S.\ Department of Energy, Office of Science, Office of Basic Energy Sciences, under Award Number DE-SC0022311. Work at Harvard was supported by the U.S.\ Department of Energy, Office of Science, Office of Basic Energy Sciences, under Award Number DE-SC0012704. Work at Argonne (nickelate sample synthesis) was supported by the U.S. DOE, Office of Science, Basic Energy Sciences, Materials Science and Engineering Division. J.Z.\ was supported by the National Natural Science Foundation of China (Grants Nos.~12074219 and 12374457). This research used resources at the SIX beamline of the National Synchrotron Light Source II, a U.S.\ DOE Office of Science User Facility operated for the DOE Office of Science by Brookhaven National Laboratory under Contract No.~DE-SC0012704.\\

\end{acknowledgments}

\section*{Data availability}
The RIXS data generated in this study have been deposited in the Zenodo database \cite{repo}.

\appendix

\section{\label{sec:synthesis}Sample synthesis}

The parent \gls*{RP} phase Pr$_4$Ni$_3$O$_{10}$ material was synthesized via the high-pressure optical floating zone method. To reduce the samples, small $c$-axis-surface-normal crystals were cleaved from the boules and heated in a flowing H$_2$/Ar gas mixture following previously reported procedures \cite{Zhang2017large}. We use tetragonal notation with space group $I4/mmm$ and lattice constants of $a=b=3.935$~\AA{}, $c=25.485$~\AA{} to describe reciprocal space. The intra-trilayer spacing $d^\prime=3.185$\AA{} and the inter-trilayer distance is $d=c/2=12.74$\AA{}. The excellent quality of the samples was confirmed by previous studies \cite{Lin2021strong, Shen2022Role, Shen2023electronic, Fabbris2023resonant}.

The \LSCO{} film used in the current study was synthesized using the atomic-layer-by-layer molecular beam epitaxy (ALL-MBE) technique~\cite{Bozovic2001atomic}. Prior studies of samples prepared via ALL-MBE have confirmed the high sample quality~\cite{Dean2012spin, Dean2013persistence,  Meyers2017doping, Shen2022Role}. The lattice constants of \LSCO{} are $a=b\approx3.76$~\AA{} and $c=13.17$~\AA{}. Layers are spaced by $d=c/26.8$\AA{}.

For both materials, momentum transfer is denoted using reciprocal lattice units (r.l.u.) notation as $\Q{}=H\bm{a}^{\ast}+K\bm{b}^{\ast}+L\bm{c}^{\ast}$. 

\section{\label{sec:RIXS}RIXS measurements}

High-energy-resolution \gls*{RIXS} measurements were performed at the SIX beamline at the \gls*{NSLS-II}. All the \gls*{RIXS} data presented here were collected with incident $\sigma$ polarization at the pre-peak of the O $K$ edge, where the plasmon intensities are maximized (Ref.~\cite{supp} Sec.~S1). During the experiments, we put the crystalline ($H$, 0, 0) and (0, 0, $L$) directions in the horizontal scattering plane [Fig.~\ref{fig:Hdisp}(a)]. The spectra were collected with an energy resolution of around 30~meV. A multi-peak fitting procedure was employed to extract the plasmon parameters, including the excitation energies and peak widths (Ref.~\cite{supp} Sec.~S2).

\section{\label{sec:RPA1}RPA calculations: Single layer model}

In the single layer \gls*{RPA} calculations, we treat the electronic dispersion as ${\epsilon}_{\bm{k}} = {\epsilon}^{\parallel}_{\bm{k}} + {\epsilon}^{\perp}_{\bm{k}}$, where 
\begin{equation}\label{eq:eps_par}
\begin{split}
{\epsilon}^{\parallel}_{\bm{k}} =& -2t\left[\cos(k_x a)+\cos(k_y a)\right] \\
&-4t^\prime\cos(k_x a)\cos(k_y a)-\mu 
\end{split}
\end{equation}
and 
\begin{equation}\label{eq:eps_perp}
{\epsilon}^{\perp}_{\bm{k}} = -2t_z\left[\cos(k_x a)-\cos(k_y a)\right]^2\cos(k_z d)
\end{equation}
are the dispersions parallel and perpendicular to the $ab$ plane, respectively \cite{Hepting2023Evolution}. $k_x$ and $k_y$ denote the in-plane momenta, expressed in units of $2\pi / a$, while $k_z$ represents the out-of-plane momentum in units of $2\pi / d$. Here, $a$ is the in-plane lattice constant and $d$ is defined in Fig.~\ref{fig:Hdisp}(a). The chemical potential $\mu$ is set by imposing $1-\delta = \frac{2}{N}\sum_{\bm{k}} n_F(\epsilon_{\bm{k}})$, where $\delta$ is doping and 
$n_\text{F}(\epsilon_{\bm{k}}) = 1/\left[e^{(\epsilon_{\bm{k}} - \mu)\beta} + 1\right]$
is the Fermi function, $\beta = 1/k_\text{B}T$ is the inverse temperature, and $N$ is the total number of sites \cite{Nag2020detection}. Note that all the calculations were performed at $T=40$~K.

In the single-layer case, the \gls*{RPA} charge susceptibility on the Matsubara frequency axis is given by
\begin{equation}
\chi({\boldsymbol{q}},\mathrm{i}\omega_m) = \frac{\chi_0({\boldsymbol{q}},\mathrm{i}\omega_m)}{1-V(\bm{q})\chi_0({\boldsymbol{q}},\mathrm{i}\omega_m)}, 
\end{equation}
where $\omega_m = 2m\pi/\beta$ is a bosonic Matsubara frequency,  
\begin{equation*}
\chi_0({\boldsymbol{q}},\omega_m) = \frac{2}{N\beta}\sum_{\boldsymbol{k},n}{G_0({\boldsymbol{k}}+{\boldsymbol{q}},\mathrm{i}\omega_n+\mathrm{i}\omega_m)G_0({\boldsymbol{k}},\mathrm{i}\omega_n)} 
\end{equation*}
is the Lindhard susceptibility, $G_0(\boldsymbol{k},\mathrm{i}\omega_n) = 1/(\mathrm{i}\omega_n - \epsilon_{\boldsymbol{k}})$ is the noninteracting Green's function, $\omega_n = (2n+1)\pi/\beta$ is a fermionic Matsubara frequency, 
and 
\begin{equation}
V(\bm{q}) = \frac{V_c}{\alpha[2-\cos{(q_xa)}-\cos{(q_ya)}]+1-\cos{(q_zd)}}
\end{equation}
is the long-range Coulomb interaction for a layered electron gas~\cite{Nag2020detection}. Here, $\alpha$ describes the anisotropy. 

The imaginary part of the charge susceptibility on the real axis $\chi^{\prime\prime}({\boldsymbol{q}},\omega)$, which is the quantity we wish to compare to our measurements, can be obtained by analytical continuation $\chi(\boldsymbol{q},\mathrm{i}\omega_n \rightarrow \omega+\mathrm{i\gamma})$. In this case, the Lindhard function on the real frequency axis is given by 
\begin{equation}
    \chi_0(\boldsymbol{q},\omega+\mathrm{i}\gamma)= -\frac{2}{N}\sum_{\boldsymbol{k}}{\frac{n_{\mathrm{F}}({\boldsymbol{k}}+{\boldsymbol{q}})-n_{\mathrm{F}}({\boldsymbol{k}})}{\omega_n-\epsilon_{{\boldsymbol{k}}+{\boldsymbol{q}}}+\epsilon_{\boldsymbol{k}}+\mathrm{i}\gamma}}. 
\end{equation}

In our numerical calculations, we take $\gamma = 5$ meV to account for resolution broadening of the charge susceptibility. We further fix 
$\alpha = 3.5$~\cite{Nag2020detection} and $t^{\prime}/t=-0.3$ based on established theoretical and experimental literature, which makes $t$, $t_z/t$ and $V_c/t$ the only free parameters. Furthermore, since $t$ dictates the overall energy scale and is accurately constrained by the experimental plasmon energy, we tuned $t_z/t$ and $V_c/t$ to search for the best combination of parameters that can reproduce the \LSCO{} ($\delta=0.35$) plasmon dispersions (Ref.~\cite{supp}, Fig.~S4). The best fit parameters are given in Table~\ref{table:Parameters}. Note that the change in plasmon dispersion resulting from reducing $V_c$ can be partly compensated by increasing $t$, so we further use \gls*{ARPES} data to constrain the latter, obtaining a unique solution.
 
\section{\label{sec:RPA2}RPA calculations: Trilayer model}

In the trilayer model, we use $d=3.286a$ and $d^\prime=0.8215a$ to denote the out-of-plane inter- and intra-trilayer distances, respectively [see Fig.~\ref{fig:Hdisp}(a)]. As such the out-of-plane momentum is reported in units of $2\pi/d$. 

We model the noninteracting electronic structure of the multilayer crystal using the Hamiltonian $H = \sum_{\boldsymbol{k},\sigma} \Psi^\dagger_{\boldsymbol{k},\sigma}\hat{h}_{\boldsymbol{k}}\Psi^{\phantom\dagger}_{\boldsymbol{k},\sigma}$, 
where $\Psi^{\dagger}_{\boldsymbol{k},\sigma} = [c^\dagger_{\boldsymbol{k},1,\sigma}, c^\dagger_{\boldsymbol{k},2,\sigma}, c^\dagger_{\boldsymbol{k},2,\sigma}]$ is a row vector of operators $c^\dagger_{\boldsymbol{k},\alpha,\sigma}$ are creation operators for a spin $\sigma$ electron in layer $\alpha$ and 
\begin{align}\label{eq:hk_mat}
\hat{h}_{\boldsymbol{k}} &= \begin{pmatrix}
  {\epsilon}^{\parallel}_{\bm{k}}+\frac{\Delta\mu}{3} & {\epsilon}^{\perp}_{\bm{k}} & 0 \\ 
  {\epsilon}^{\perp*}_{\bm{k}} & {\epsilon}^{\parallel}_{\bm{k}}-\frac{2\Delta\mu}{3} & {\epsilon}^{\perp}_{\bm{k}} \\
  0 & {\epsilon}^{\perp*}_{\bm{k}} & {\epsilon}^{\parallel}_{\bm{k}}+\frac{\Delta\mu}{3}
\end{pmatrix}.
\end{align}
Here, $\Delta\mu$ is the potential difference between layers, which is introduced to account for inequivalent doping levels in the inner ($n = 2$) and outer ($n = 1,3$) layers, as observed in \gls*{ARPES} measurements on \Pr438{}~\cite{Li2023Electronic}. The in-plane component of the dispersion ${\epsilon}^{\parallel}_{\bm{k}}$ is defined as in Eq.~\eqref{eq:eps_par} while the out-of-plane component is 
\begin{equation}
{\epsilon}^{\perp}_{\bm{k}} =-2t_z[\cos{(k_xa)}-\cos{(k_ya)}]^2e^{ik_z d^\prime}.
\end{equation}
The electronic dispersions in the band basis is obtained by diagonalizing $\hat{\epsilon}_{\boldsymbol{k}}= U^\dagger_{\bf k}\hat{h}^{\phantom\dagger}_{\bf k}U^{\phantom\dagger}_{\bf k}$, 
where $\hat{\epsilon}_{\bf k}$ is a diagonal matrix representing the eigenvalues of $\hat{h}_{\boldsymbol{k}}$. 

The \gls*{RPA} susceptibility, written in the orbitals basis, is a $3\times 3$ matrix 
given by 
\begin{equation}
    \chi({\boldsymbol{q}},\mathrm{i}\omega_m) =  \left[\mathbb{I}-V(\bm{q})\chi_0({\boldsymbol{q}},\mathrm{i}\omega_m)\right]^{-1}\chi_0({\boldsymbol{q}},\mathrm{i}\omega_m).
\end{equation}
Here, $\mathbb{I}$ is an identity matrix, $V({\boldsymbol{q}})$ is the long-range Coulomb interaction for coupled sets of layered gases, and 
$\chi_0(\boldsymbol{q},\mathrm{i}\omega_m)$ is the Lindhard function, written in the layer basis. The Coulomb interaction is defined as~\cite{Nakata2025Outofphasea} 
\begin{align}
V({\boldsymbol{q}}) &= \begin{pmatrix}
  V_{11} & V_{12} & V_{13} \\ 
  V^*_{12} & V_{11} & V_{12} \\
  V^*_{13} & V^*_{12} & V_{11}
\end{pmatrix},
\end{align}
where
\begin{widetext}
\begin{equation}
\begin{split}
V_{11}(\bm{q}) &=\frac{V_c}{\bm{q}_\parallel}\left(\frac{\sinh(|\bm{q}_\parallel| d)}{\cosh(|\bm{q}_\parallel|d) - \cos(q_z d)}\right) \\
V_{12}(\bm{q}) &=\frac{V_c}{\bm{q}_\parallel}\left(\frac{\sinh(|\bm{q}_\parallel|(d-d^\prime))+e^{-\mathrm{i}q_zd}\sinh(|\bm{q}_\parallel| d^\prime)}{\cosh(|\bm{q}_\parallel|d) - \cos(q_zd))}\right)e^{-\mathrm{i}q_zd^\prime} \\
V_{13}(\bm{q}) &=\frac{V_c}{\bm{q}_\parallel}\left(\frac{\sinh(|\bm{q}_\parallel|(d-2d^\prime))+e^{-\mathrm{i}q_zd}\sinh(|\bm{q}_\parallel| 2d^\prime)}{\cosh(|\bm{q}_\parallel|d) - \cos(q_zd))}\right)e^{-\mathrm{i}2q_zd^\prime} .
\end{split}
\end{equation}
The \gls*{RIXS} response is obtained from the imaginary part of the analytically continued \gls*{RPA} susceptibility~\cite{Bejas2024Plasmon} 
\begin{equation}
    \mathrm{Im}\chi_{c}(\boldsymbol{q},\omega) = -\sum_{\alpha,\beta} \mathrm{Im}\chi_{\alpha,\beta}(\boldsymbol{q},\mathrm{i}\omega_m \rightarrow \omega + \mathrm{i}\gamma). 
\end{equation}
The Lindhard function in this case is given by
\begin{align*}
\chi^0_{\alpha,\beta}(\boldsymbol{q},\omega) = -\frac{2}{N}\sum_{{\boldsymbol{k}},\mu,\nu} \int_{-\infty}^{\infty}dx \int_{-\infty}^{\infty}dy A_{\nu, \nu}(\bm{k}+\bm{q}, y) A_{\mu, \mu}(\bm{k}, y) U_{\mu,\alpha}({\boldsymbol{k}})U^*_{\mu,\beta}({\boldsymbol{k}})U_{\nu,\beta}({\boldsymbol{k}}+{\boldsymbol{q}})U^*_{\nu,\alpha}({\boldsymbol{k}}+{\boldsymbol{q}}) \frac{n_\mathrm{F}(y)-n_\mathrm{F}(x)}{\omega - y + x + \mathrm{i}\gamma}, 
\end{align*}
where $\alpha$ and $\beta$ are orbital indices and $\mu$ and $\nu$ are band indices and 
$A(\boldsymbol{k},\omega) =-\frac{1}{\pi}\mathrm{Im}G(\boldsymbol{k}, \omega)$ is the spectral function in the band basis. 
\end{widetext}

In the noninteracting limit, the spectral function is diagonal with elements  $A_{\nu,\nu}(\boldsymbol{k},\omega) = \delta(\omega-\epsilon_\nu(\boldsymbol{k}))$, where $\epsilon_\nu(\boldsymbol{k})$ are the eigenvalues of Eq.~\eqref{eq:hk_mat}. In the interacting case, we assume that the self-energy is diagonal and has a Fermi-liquid-like dependence $\Sigma_{\nu,\nu}(\omega) = -\mathrm{i}\left(\kappa_0 + \kappa\omega^2\right)$ such that $A_{\nu,\nu}(\boldsymbol{k},\omega) = -\frac{1}{\pi}\mathrm{Im}\left[\omega - \epsilon_\nu(\boldsymbol{k}) - \Sigma_{\nu,\nu}(\omega)\right]^{-1}$. Here, $\kappa_0$ and $\kappa$ control the strength of the impurity scattering and electron-electron scattering, respectively. Note that plasmon energy involves minimal sensitivity to this choice, since the lowest energy excitations arise from electrons close to the Fermi level.

For our \Pr438{} plasmon simulations, we fixed $t^{^\prime}/t =-0.3$, as in the single layer model, and $\Delta\mu/t=-0.4$ to ensure consistency with the \gls*{ARPES} Fermi surface topology (Ref.~\cite{supp} Fig.~S6)~\cite{Li2023Electronic}. Thus, similar to the single-layer model, we adjusted $t_z/t$ and $V_c/t$ to identify the optimal combination that reproduces the experimental observations (see Ref.~\cite{supp} Fig.~S5 in which we also varied $\Delta\mu/t$ for completeness). It is also worth noting that the reduced $t$ and $V_c$ in \Pr438{} compared to \LSCO{} can be concluded directly from a qualitative consideration of the \gls*{RIXS} data, thus the simulations act to confirm and quantify a trend already evident in the raw data.

While the trilayer model appears distinct from the single-layer one, it is not fundamentally more complex; rather, it represents a straightforward extension designed to incorporate the trilayer structure. Both models employ identical theoretical procedures and methods for incorporating long-range Coulomb interactions. Furthermore, the exclusion of inter-trilayer hopping is justified by the minimal magnitude of intra-trilayer hopping (0.0084~eV). Given the larger spacing between blocks, inter-trilayer hopping is expected to be even weaker, resulting in effects on the plasmon dispersion that fall below the experimental resolution.

\bibliography{refs}

\end{document}


\title{Supplemental Material: Observation of correlated plasmons in low-valence nickelates}

\author{Y. Shen\orcidlink{0000-0003-4697-4719}}
\email[]{yshen@iphy.ac.cn}
\author{W. He\orcidlink{0000-0003-3522-3899}}
\author{J. Sears\orcidlink{0000-0001-6524-8953}}
\author{Xuefei Guo\orcidlink{0000-0003-1088-6039}}
\author{Xiangpeng Luo\orcidlink{0000-0002-5471-053X}}
\author{A. Roll\orcidlink{0009-0007-6140-0242}}  
\author{J. Li\orcidlink{0000-0002-7153-6123}}
\author{J. Pelliciari\orcidlink{0000-0003-1508-7746}}
\author{Xi He\orcidlink{0000-0001-6603-2388}}
\author{I. Bo\v{z}ovi\'{c}\orcidlink{0000-0001-6400-7461}}
\author{Junjie Zhang\orcidlink{0000-0002-5561-1330}}
\author{J. F. Mitchell\orcidlink{0000-0002-8416-6424}}
\author{V. Bisogni\orcidlink{0000-0002-7399-9930}}
\author{M. Mitrano\orcidlink{0000-0002-0102-0391}}
\author{S. Johnston\orcidlink{0000-0002-2343-0113}}
\author{M. P. M. Dean\orcidlink{0000-0001-5139-3543}}
\email[]{mdean@bnl.gov}

\date{\today}

\maketitle

\renewcommand{\thesection}{S\arabic{section}}  
\renewcommand{\thetable}{S\arabic{table}}  
\renewcommand{\thefigure}{S\arabic{figure}}

\vspace{-1cm}
\tableofcontents

\section{Resonant behaviors of plasmons}

\begin{figure*}
\includegraphics{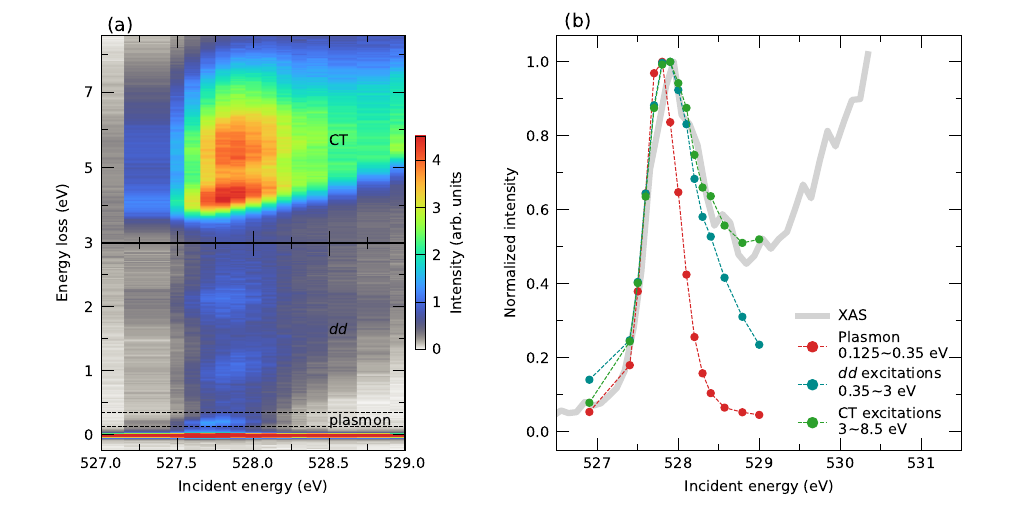}
\caption{Energy dependent \acrfull*{RIXS} spectra. (a), \gls*{RIXS} energy maps across the pre-edge regime of the O $K$ edge collected with $\theta$=22.6~$^{\circ}$ and $\sigma$ polarization. Various excitations are identified, including the charge-transfer (CT) excitations, orbital $dd$ excitations, and plasmons. Additionally, weak phonons are observed overlapping with the quasi-elastic peak. The \gls*{XAS} data were collected in partial fluorescence yield (PFY) mode. (b), Incident energy dependence of the \gls*{RIXS} spectral weights integrated in the indicated energy ranges, representing different excitations.}
\label{fig:Pr438_Edep}
\end{figure*}

\begin{figure*}
\includegraphics{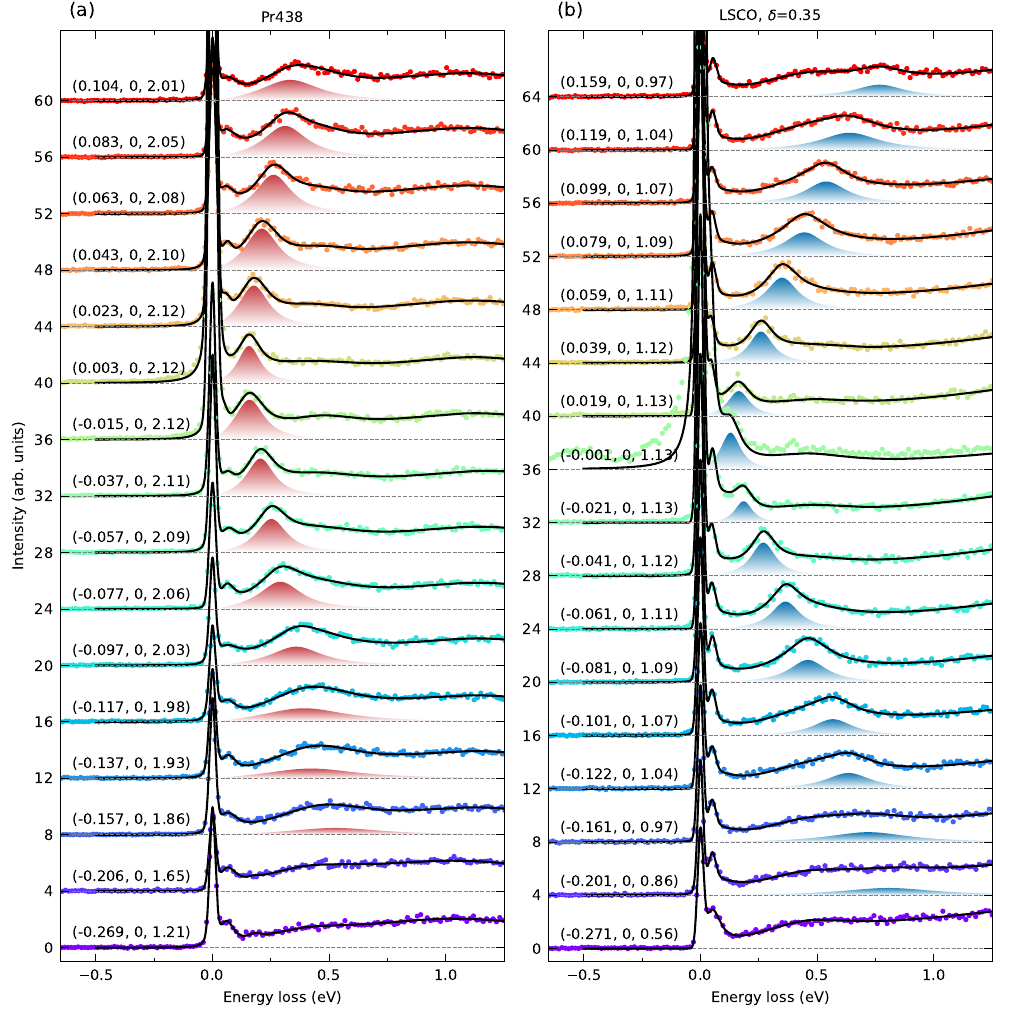}
\caption{Background subtracted O $K$-edge \gls*{RIXS} spectra for the in-plane plasmon dispersions. All the data for \Pr438{} (Pr438) and \LSCO{} (LSCO) were measured at 40~K. The dots present the data collected at the indicated \Q{} positions, and the solid lines are fitting results. The shaded areas highlight the plasmon contributions.}
\label{fig:Hdisp2}
\end{figure*}

\begin{figure*}
\includegraphics{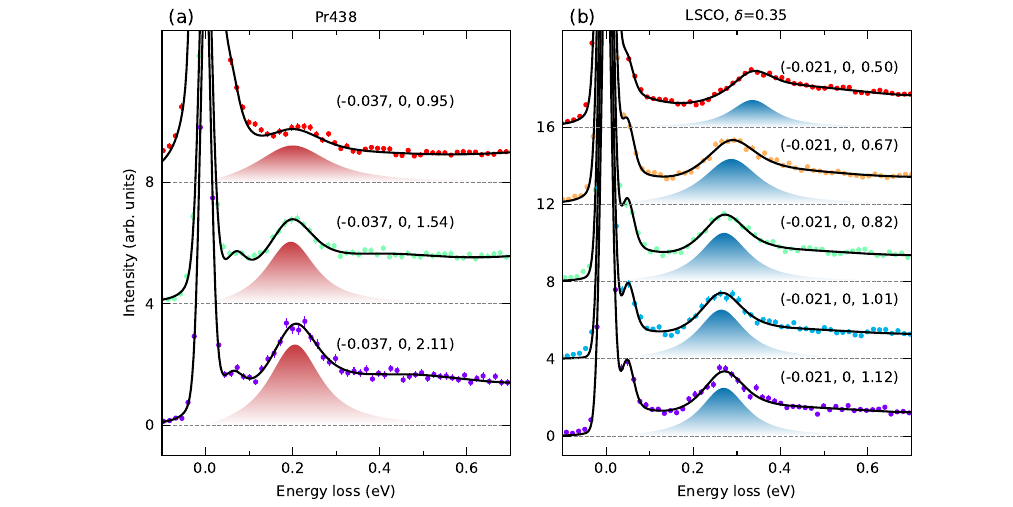}
\caption{\gls*{RIXS} spectra along the the out-of-plane direction.}
\label{fig:Ldisp2}
\end{figure*}

The overdoped low-valence nickelates $R_4$Ni$_3$O$_8$ exhibit a pre-peak feature in the O $K$-edge \gls*{XAS}~\cite{Zhang2017large}, indicative of an appreciable amount of ligand holes and nonzero Ni-O hybridization. Figure~\ref{fig:Pr438_Edep} depicts the \gls*{RIXS} energy map measured on \Pr438{} single crystals, which shows multiple Raman-like features resonant at the pre-peak regime, indicating various excitations. As revealed in our prior studies, the high-energy strong features correspond to charge-transfer excitations, and the lower-energy weak features originate from $dd$ excitations through Ni-O hybridization. Plasmons are observed with lower energy loss ($\sim$0.2~eV), which is absent in the insulating phase of \La438{}~\cite{Shen2022Role}. Its resonant behavior highlights the localized characters.

\section{Fitting of the RIXS spectra}

Figure~\ref{fig:Hdisp2} and Fig.~\ref{fig:Ldisp2} present the \gls*{RIXS} spectra for plasmon dispersions along the in-plane and out-of-plane directions, respectively. A constant background has been subtracted for all the spectra using the energy-gain side. For \Pr438{}, each spectrum was fit with five components, including a pseudo-Voigt profile, representing the quasi-elastic peak, and four damped harmonic oscillators convoluted with resolution functions determined by the quasi-elastic peak to account for the phonons at $\sim$0.08~eV, dispersive plasmons indicated by the shaded areas, and two high-energy $dd$ excitation modes. For \LSCO{}, we consider two phonon modes, one dispersionless bimagnon branch, and a tail from higher-energy charge excitations.

\begin{figure*}
\includegraphics{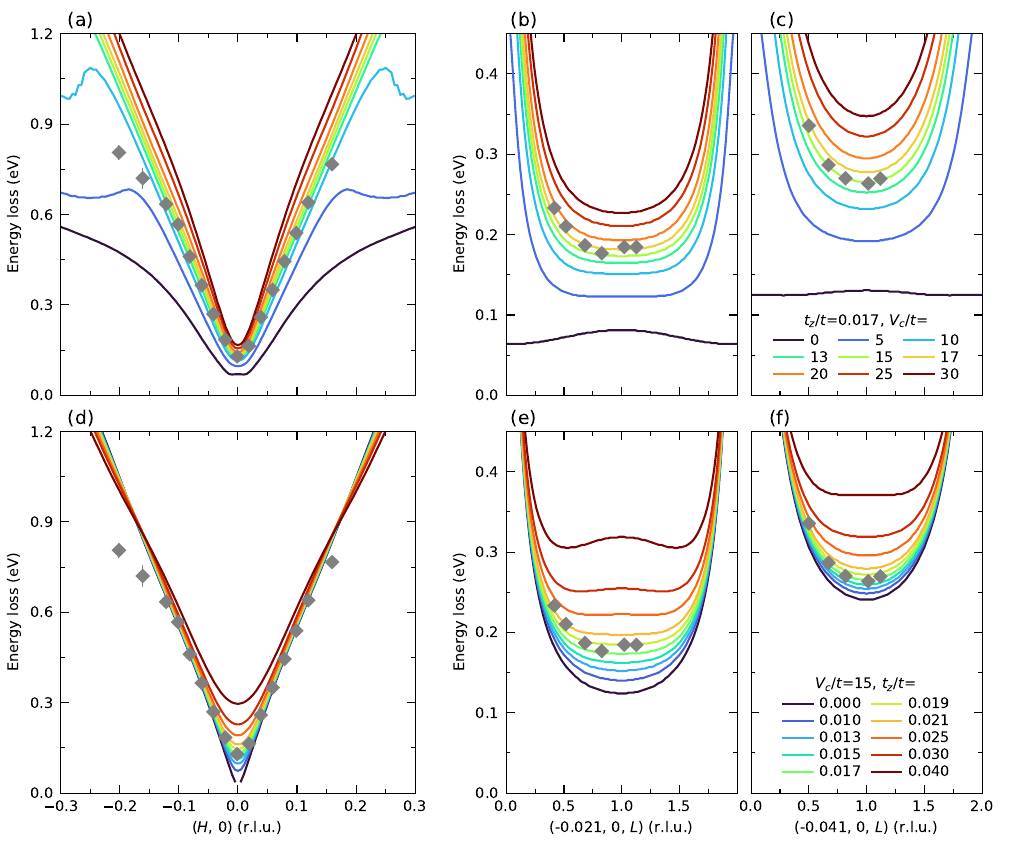}
\caption{\Acrfull*{RPA} calculations of \LSCO{} ($\delta=0.35$) plasmons as a function of tuning parameters. The diamonds are data points and the solid lines are calculation results. Based on the simulation results, the plasmon velocity for $V_c/t=0$ [dark blue line in panel (a)] is approximately 1.8~eV\ \AA{}.
}
\label{fig:LSCO_calc}
\end{figure*}

\begin{figure*}
\includegraphics{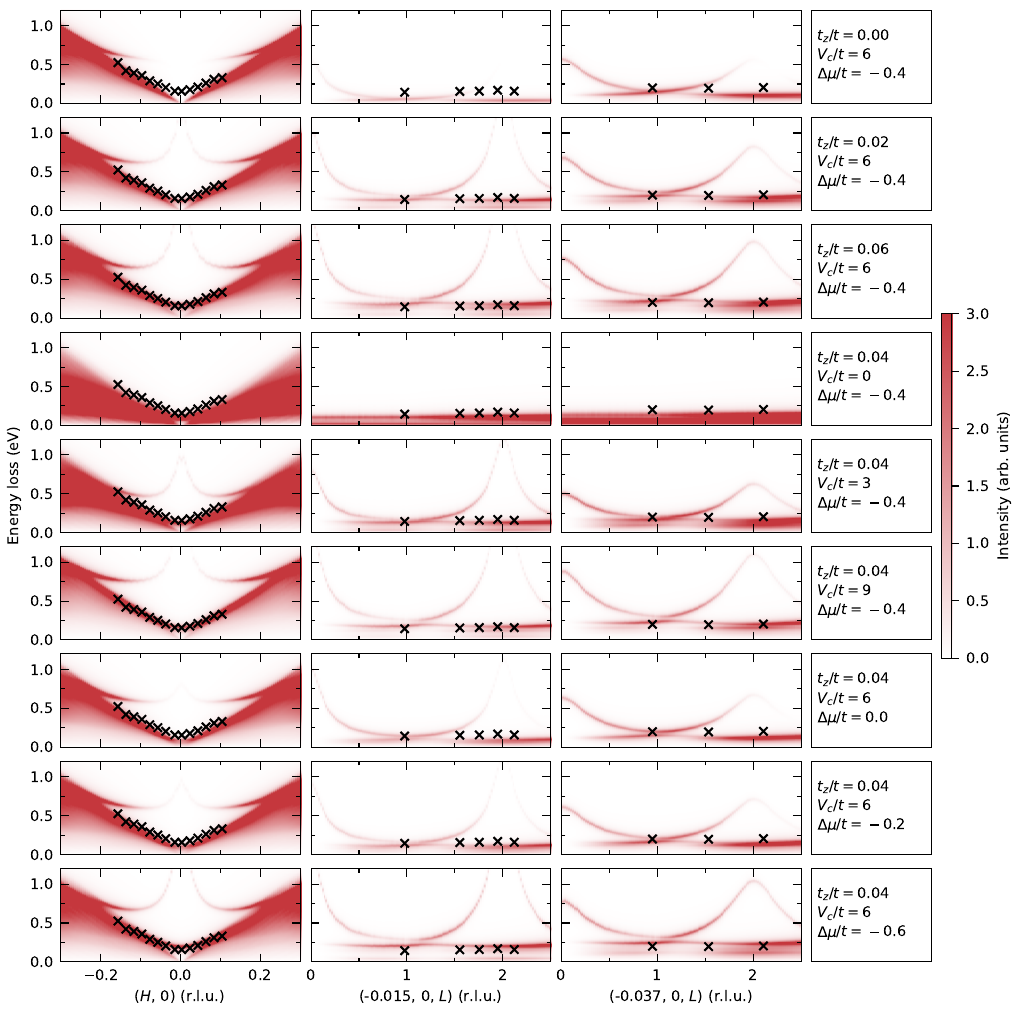}
\caption{\gls*{RPA} calculations of \Pr438{} plasmons with tuning parameters. The crossings are data points.}
\label{fig:Pr438_calc2}
\end{figure*}

\begin{figure*}
\includegraphics{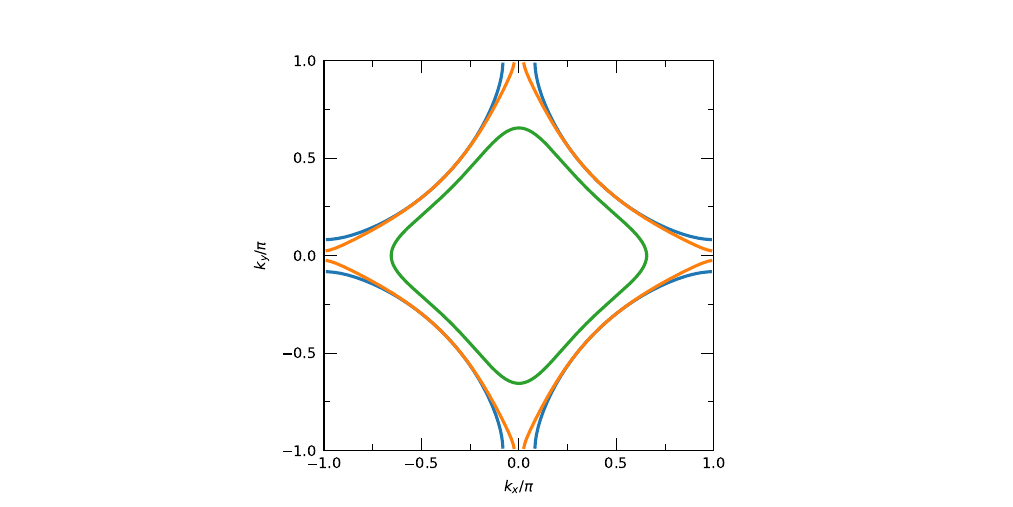}
\caption{Calculated \Pr438{} Fermi surface using the parameters in the main text.}
\label{fig:Pr438_calc2}
\end{figure*}

\begin{figure*}
\includegraphics{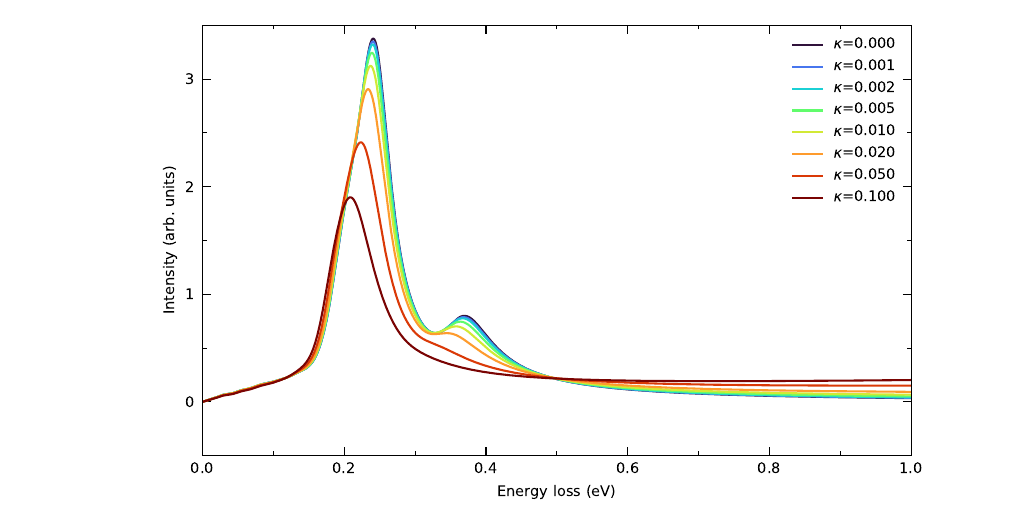}
\caption{\gls*{RPA} calculations of \Pr438{} plasmons with various self-energy broadening.}
\label{fig:Pr438_calc2}
\end{figure*}

\clearpage
\bibliography{refs}